\documentclass{PoS}

\title{Hadron gas with repulsive mean field}

\ShortTitle{Hadron gas with repulsive mean field}

\author{\speaker{Pasi Huovinen}\\ %\thanks{A footnote may follow.}\\
        Institute of Theoretical Physics, University of Wroclaw, 50204 Wroc\l aw, Poland\\
        E-mail: \email{pasi.huovinen@ift.uni.wroc.pl}}

\author{Peter Petreczky\\
        Physics Department, Brookhaven National Laboratory, Upton, NY 11973, USA\\
        E-mail: \email{petreczk@quark.phy.bnl.gov}}

\abstract{We study the QCD equation of state, and fluctuations of
  baryon number and strangeness using the hadron resonance gas model
  with repulsive mean field. We find that including both the predicted
  but not observed resonances, a.k.a. missing states, and the
  repulsive mean field into the resonance gas model leads to better
  description of the lattice results. The repulsive mean field is
  particularly important for the higher order baryon number
  fluctuations. }

\FullConference{XIII Quark Confinement and the Hadron Spectrum - Confinement2018\\
		31 July - 6 August 2018\\
		Maynooth University, Ireland}

\begin{document}

\section{Introduction}

Below the chiral crossover at zero net baryon density QCD
thermodynamics can be well approximated by the hadron resonance gas
(HRG) model. This model is based on the idea that the effect of
interactions in hadron gas can be mimicked by including resonances as
additional free particles in hadron gas---an idea, which can be
justified using the relativistic virial expansion based on the
S-matrix approach~\cite{Dashen:1969ep}. In most interactions, after
summation over spin and isospin channels, the second virial
coefficient is dominated by the resonance contribution, and thus the
interacting gas of hadrons can be approximated as a gas of
non-interacting hadrons and hadronic
resonances~\cite{Venugopalan:1992hy}. The validity of the HRG model is
further corroborated by the equation of state obtained in lattice QCD
calculations, which agrees well with the HRG equation of state (see
e.g.~Refs.~\cite{Borsanyi:2010bp,Borsanyi:2013bia,Bazavov:2014pvz,
  Bazavov:2017dsy}). The fluctuations and correlations of conserved
charges, defined as derivatives of pressure with respect to chemical
potentials, have also been studied in the HRG model (see
e.g.~Ref.~\cite{Huovinen:2009yb}).

However, the applicability of HRG must have its limits. The
cancellation of non-resonant contributions to thermodynamics is
somewhat accidental~\cite{Fernandez-Ramirez:2018vzu} and does not
happen for all thermodynamic quantities~\cite{Lo:2017lym}. There are
also hadronic interactions which are not dominated by resonances,
e.g.~there are no resonances in nucleon-nucleon scatterings. Last but
not least, the HRG model is based on free particle properties, but
close to the chiral transition there are in-medium modifications to
hadron properties~\cite{Kelly:2018hsi,Aarts:2017rrl}. Therefore, it is
important to check the HRG model by comparing it to recent lattice QCD
calculations of different thermodynamic quantities.

It is known that for higher order ($>2$) fluctuations the agreement
between lattice and HRG is not good for temperatures close to the
chiral transition. It has been argued that this is not a signal of
approaching transition, but rather due to repulsive baryon-baryon
interactions which the HRG model does not describe~\cite{Albright:2015uua,
  Vovchenko:2016rkn,Vovchenko:2017zpj}. If that is the case, the
repulsive baryon-baryon interactions must be included in the model
when HRG is extended to non-zero baryon densities, since the larger the
baryon density, the larger the effect of repulsive interactions. As a
step towards finding a convenient description of baryon repulsion, we
study the trace anomaly and fluctuations of baryon number and
strangeness using HRG with repulsive mean field.

\section{Nucleon gas with repulsive mean field and virial expansion}

\begin{figure}
  \begin{minipage}[t]{70mm}
    \includegraphics[width=75mm]{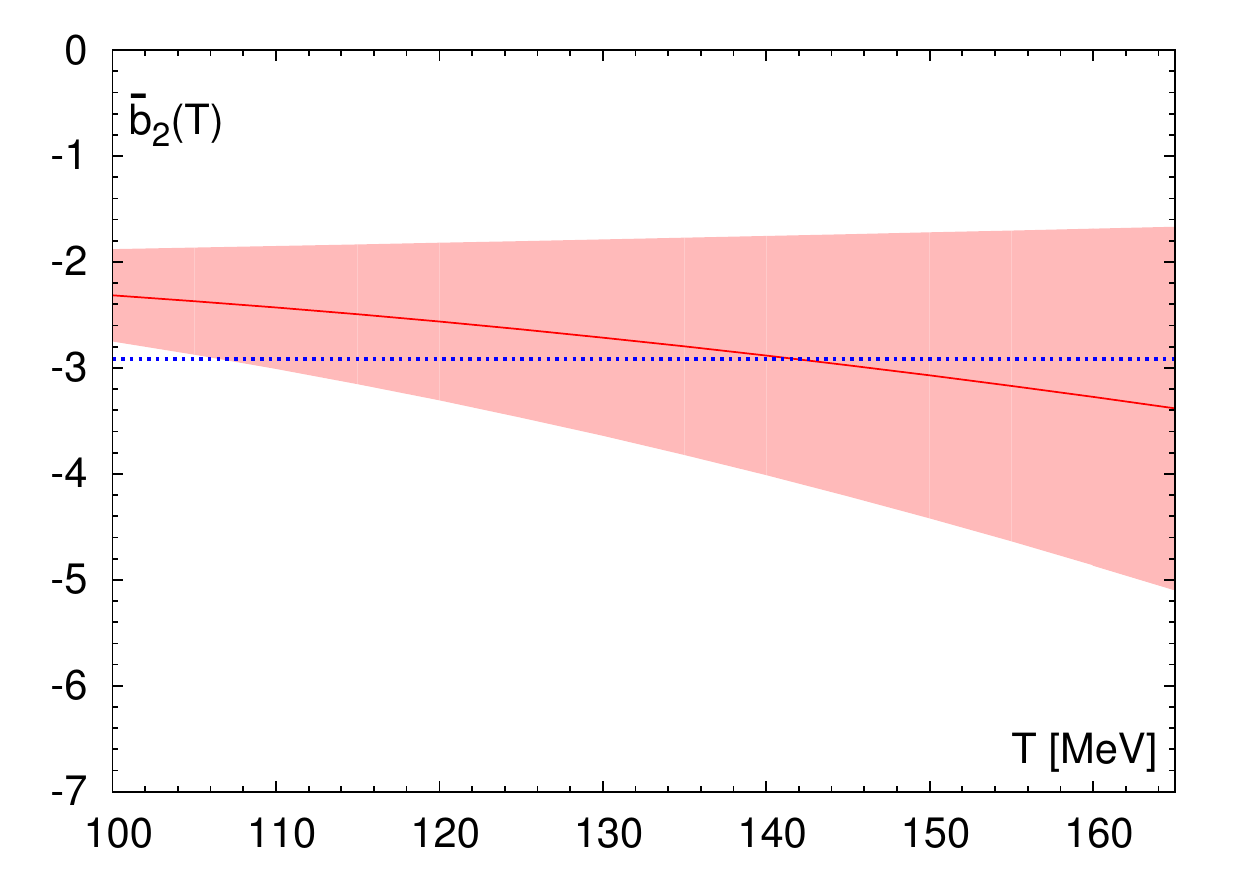}
      \caption{The reduced second virial coefficient $\bar b_2$ as
        function of temperature (red) , and the corresponding coefficient for
        mean field parameter $K = 250\,{\rm MeV/fm}^3$ (blue).}
      \label{fig:b2}
  \end{minipage}
    \hfill
  \begin{minipage}{70mm}
    \includegraphics[width=85mm]{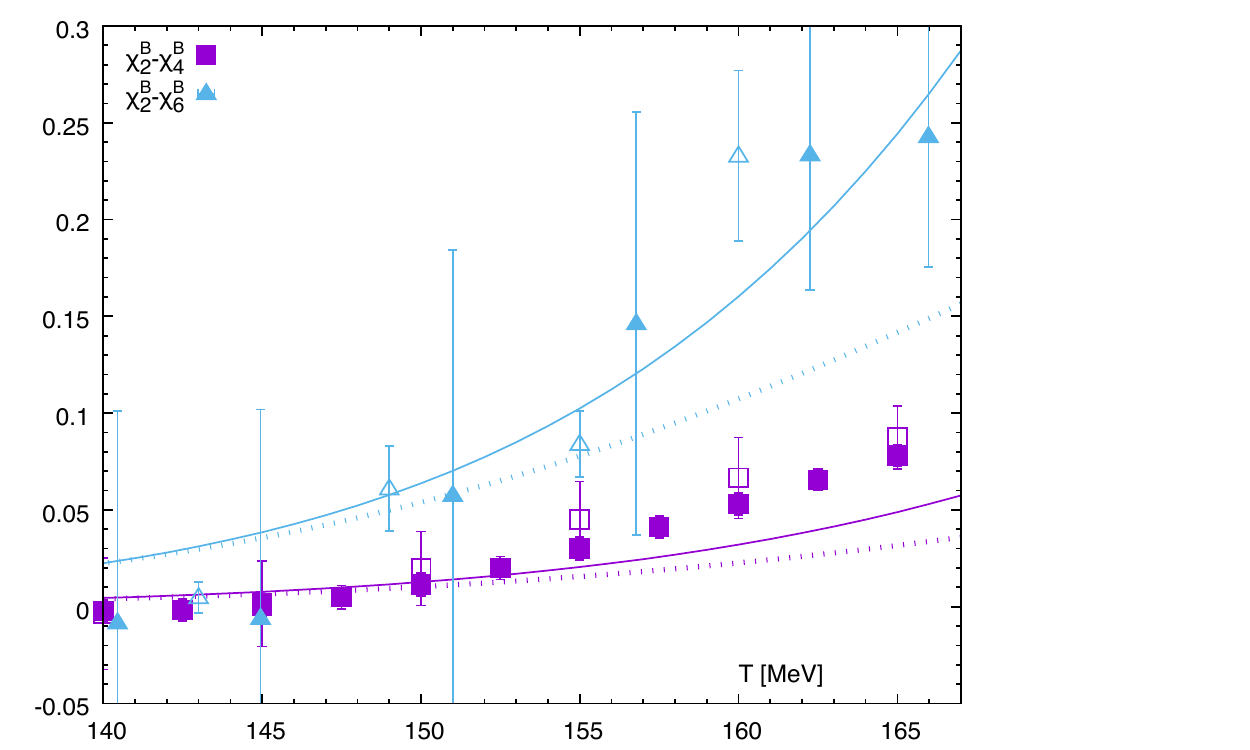}
      \caption{The difference between fourth and second order baryon
        number fluctuations (squares) and the the sixth and second
        order baryon number fluctuations (triangles). The dashed lines
        correspond to the exact mean-field calculations, while the
        solid line to the expanded mean field (see text). The filled
        symbols correspond to lattice results of
        Ref.~\cite{Bazavov:2017dus}. The open symbols are the lattice
        data for $\chi_4^B$ from Ref.~\cite{Borsanyi:2014ewa} and
        $\chi_6^B$ from Ref.~\cite{DElia:2016jqh}, respectively.}
      \label{fig:old}
  \end{minipage}
\end{figure}

According to the virial expansion, pressure of the interacting
nucleon gas can be written as~\cite{Huovinen:2017ogf}
\begin{equation}
p(T,\mu)=p_0(T)\cosh(\beta \mu)+2 b_2(T) T \cosh(2 \beta \mu),~\beta=1/T,
\end{equation}
where 
\begin{equation}
p_0(T)=\frac{4 M^2 T^2}{\pi^2} K_2(\beta M)
\end{equation}
is the pressure of free nucleon gas at zero chemical potential, and
$b_2(T)$ is the second virial coefficient. The second virial
coefficient can be written as
\begin{equation}
  b_2(T) = \frac{2 T}{\pi^3}
           \int_0^{\infty} dE (\frac{ME}{2}+M^2) K_2\left(2 \beta \sqrt{\frac{M E}{2}+M^2}\right)
           \frac{1}{4 i}{\rm Tr} \left[ S^{\dagger}\frac{dS}{dE}-\frac{dS^{\dagger}}{dE} S \right],
\end{equation}
where $S$ is the scattering S-matrix and $E$ the kinetic energy in the
lab frame. Furthermore, $M$ is the nucleon mass and $K_2(x)$ is the
Bessel function of second kind. The virial coefficient $b_2(T)$ can be
evaluated using the experimentally measured phase shifts to
parametrise the $S$-matrix, and it turns out that $b_2(T)$ is
negative~\cite{Huovinen:2017ogf}, see Fig.~\ref{fig:b2}.

For the comparison with the mean-field approach it is convenient to write
the pressure as
\begin{equation}
p(T,\mu)=p_0(T) ( \cosh(\beta \mu)+\bar b_2(T) K_2(\beta M) \cosh(2 \beta \mu) ),
\label{red_vir}
\end{equation}
where
\begin{equation}
\bar b_2(T)=\frac{2 T b_2(T)}{p_0(T) K_2(\beta M)}.
\end{equation}
is the reduced virial coefficient. 

In the mean-field approach pressure can be written in Boltzmann
approximation as~\cite{Huovinen:2017ogf}
\begin{equation}
  p(T,\mu)=T(n_b+\bar n_b)+\frac{K}{2} (n_b^2+\bar n_b^2),
\end{equation}
where $n_b$ and $\bar n_b$ are the densities of nucleons and anti-nucleons,
respectively, defined by the following self-consistent relations:
\begin{equation}
       n_b=4 \int \frac{d^3 p}{(2 \pi)^3} e^{-\beta(E_p-\mu+U)},~~
  \bar n_b=4 \int \frac{d^3 p}{(2 \pi)^3} e^{-\beta(E_p+\mu+\bar U)},~E_p^2=p^2+M^2.
\label{densities}
\end{equation}
Here $U=K n_b$ and $\bar U=K \bar n_b$ are the mean-field potentials for 
nucleons and anti-nucleons. The chemical potential corresponding to the net
nucleon density is denoted by $\mu$.
This form of the pressure ensures thermodynamic consistency,
i.e.~$\partial p/\partial \mu=n_b-\bar n_b$.
Since we are mostly interested in the region of not too high baryon densities
we expand the above expressions in $\beta U=\beta K n_b$ and $\bar U=K \bar n_b$, and
keep only the leading order terms in $K$. This simplifies the pressure to
\begin{equation}
  p(T,\mu) = T(n_b^0+\bar n_b^0)-\frac{K}{2} \left(\left(n_b^0\right)^2
            +\left(\bar n_b^0\right)^2 \right),
 \label{expand}
\end{equation}
where $n_b^0$ ($\bar n_b^0$) is the free nucleon (anti-nucleon) density.
Equation~(\ref{expand}) can also be written as
\begin{equation}
  p(T,\mu) = \frac{4 T^2 M^2}{\pi^2} K_2(\beta M) \cosh(\beta \mu)
                                   -4 K \frac{T^2 M^4}{\pi^4} K_2^2(\beta M)\cosh(2 \beta \mu),
\end{equation}
which is very similar to the virial expansion pressure
Eq.~(\ref{red_vir}).  By comparing these two results, one can
determine the value of the mean field coefficient $K$ in a limited
temperature range. Since $\bar b_2(T)$ turns out to be negative,
$K>0$, and the mean field is repulsive. The temperature dependence of
$\bar b_2$ is shown in Fig.~\ref{fig:b2} and it turns out to be
relatively mild except maybe at the highest temperature. The value of
$\bar b_2$ is consistent with $K=250\,{\rm MeV/fm}^3$. The largest
value allowed for $K$ by the virial expansion is around $K=450\,{\rm MeV/fm}^3$.

It is straightforward to generalise the mean-field approach to a
multicomponent system if one assumes that the repulsive mean-field is
the same for all ground state baryons~\cite{Huovinen:2017ogf}, and the
baryon resonances are not affected by the mean field. Within this
approach we calculated baryon number fluctuations defined as
derivatives of the pressure with respect to baryon chemical
potential~\cite{Huovinen:2017ogf}
\begin{equation}
  \chi_n^B = T^n \frac{\partial^n p/T^4}{\partial \mu_B^n}.
\end{equation}
In Fig.~\ref{fig:old} we show the differences of the baryon number
fluctuations, $\chi_4^B-\chi_2^B$ and $\chi_6^B-\chi_2^B$ obtained in
HRG model with the mean field and the value of
$K=450\,{\rm MeV/fm}^3$. As seen, the model can qualitatively explain
these differences, but at quantitative level it underpredicts the
lattice data even if the value of $K$ is the largest the experimental
data allows. We note that at temperatures above $150$ MeV the expanded
mean-field expression (\ref{expand}) is no longer accurate and one
should use the exact mean-field expressions with particle densities
determined self-consistently,
Eq.~(\ref{densities})~\cite{Huovinen:2017ogf}. The exact mean-field
results are shown as dashed lines in the figure.

\section{Comparison with lattice QCD}

\begin{figure}
  \begin{minipage}[t]{70mm}
    \includegraphics[width=75mm]{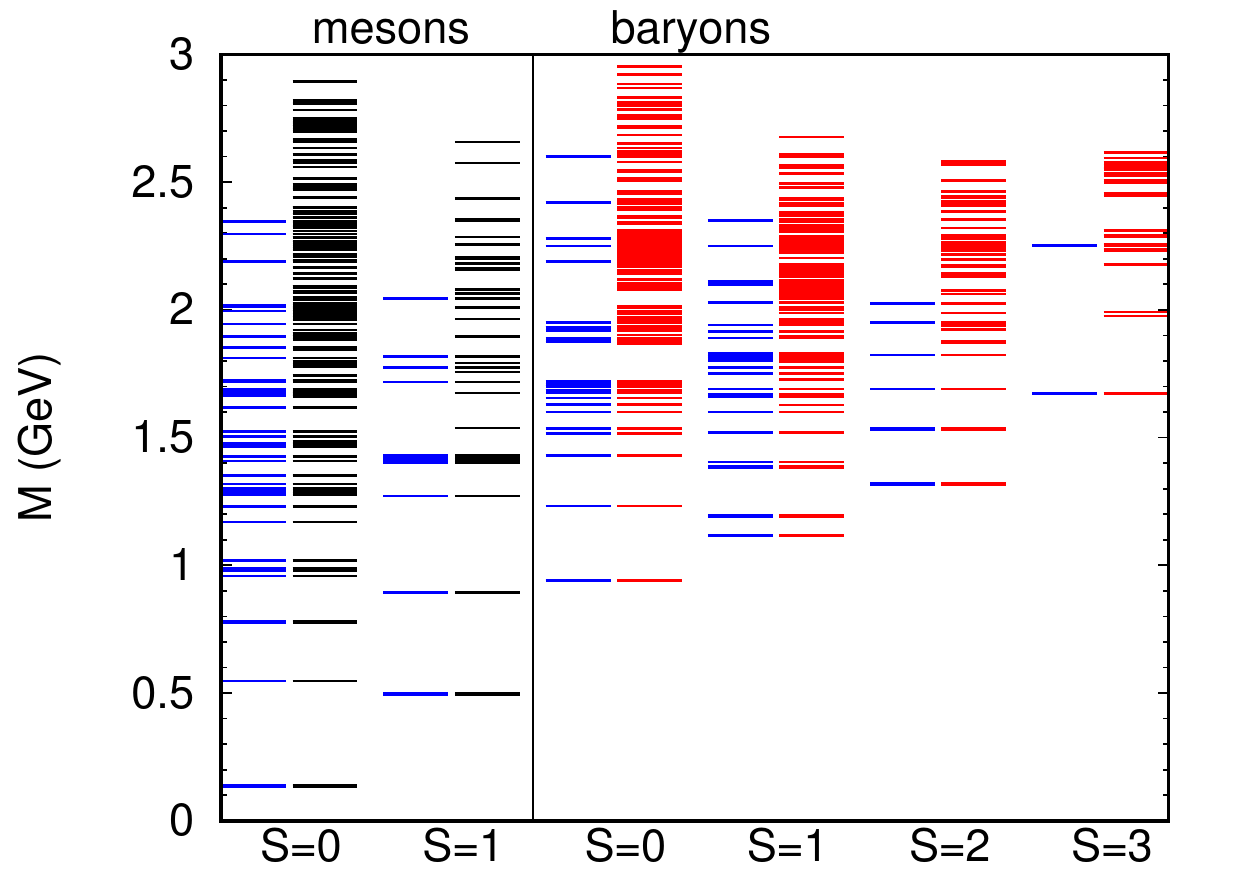}
    \caption{The mass spectrum of non-strange and strange mesons,
      non-strange baryons, and baryons of strangeness $S=1$, 2 and 3
      as given in Particle Data Group's 2016 summary tables
      (blue)~\cite{PDG}, or by PDG but augmented by additional states
      from quark model calculations from Ref.~\cite{Ebert:2009ub} (mesons,black)
      and Refs.~\cite{Loring:2001kx,Loring:2001ky} (baryons, red)}
    \label{fig:states}
  \end{minipage}
    \hfill
  \begin{minipage}[t]{70mm} \hspace*{-5mm}
    \includegraphics[width=80mm]{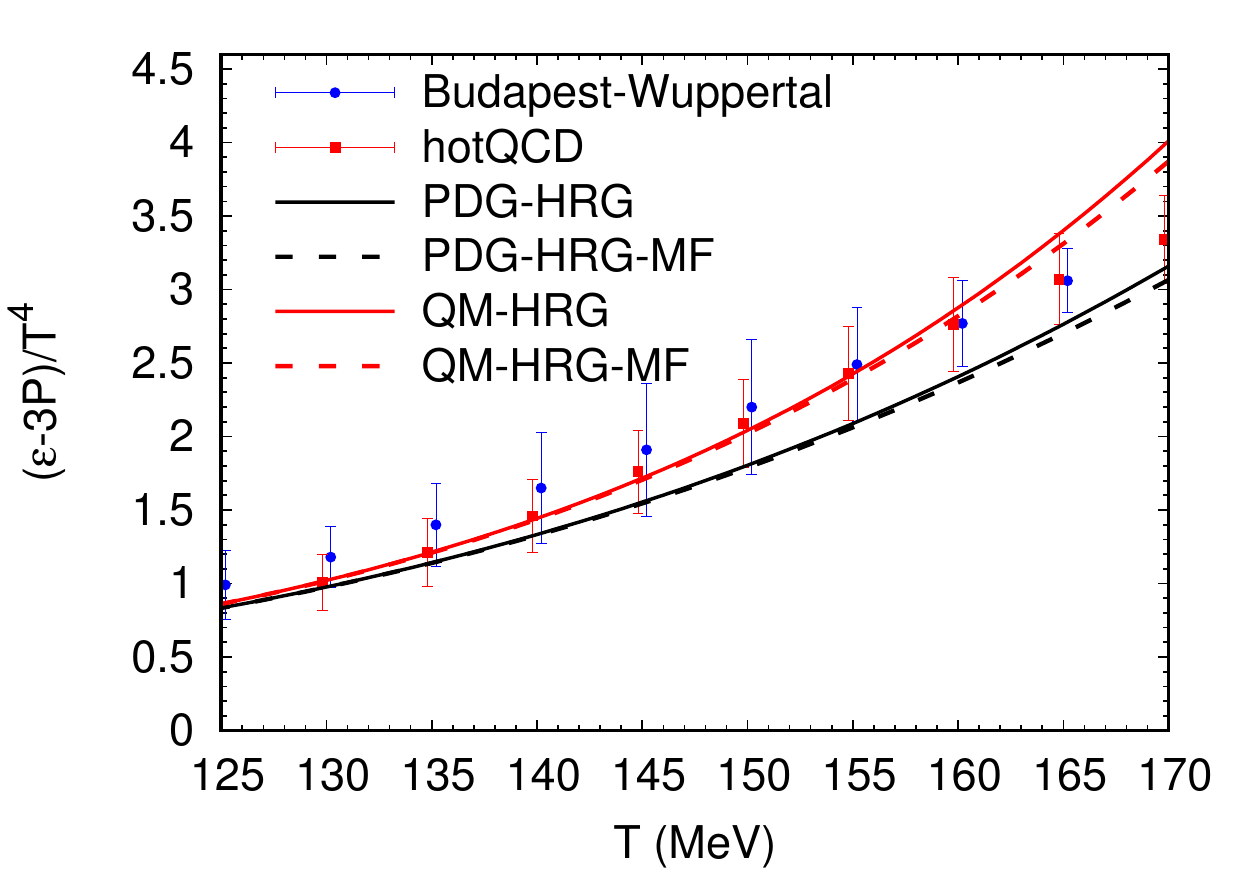}
      \caption{The trace anomaly in HRG model with the Particle Data
        Group (PDG-HRG) and Quark Model (QM-HRG) lists of resonances
        compared to the lattice results
        \cite{Borsanyi:2013bia,Bazavov:2014pvz}. The dashed lines
        correspond to HRG model with repulsive mean field.}
    \label{fig:trace}
  \end{minipage}
\end{figure}

After fixing the value of the mean field parameter, we compare the HRG
calculations with repulsive mean field with lattice QCD results on the
trace anomaly, fluctuations of baryon number up to sixth order, and
second order strangeness fluctuations. First we note that there are
many baryon resonances which are not experimentally observed but
predicted by quark models and lattice QCD, so-called missing states.
It was found that inclusion of these states in the HRG description
improves the description of the strangeness fluctuations and
baryon-strangeness correlation~\cite{Bazavov:2014xya}. Therefore, in
our analysis we will include the missing baryons from the quark model
calculations of Refs.~\cite{Loring:2001kx,Loring:2001ky}, missing
mesons from Ref.~\cite{Ebert:2009ub}, and label the corresponding
results as QM-HRG, whereas the HRG based on the Particle Data Group's
(PDG) resonance list~\cite{PDG} is labeled PDG-HRG. The difference in
the mass spectrum of these two approaches is depicted in
Fig.~\ref{fig:states}.

In our previous analysis we assumed that the baryon resonances are not
affected by the mean field~\cite{Huovinen:2017ogf}. This is not
realistic because if the density of ground state baryons is reduced
there will be also fewer baryon resonances in the system. Therefore,
in the present analysis we include the effect of the repulsive mean
field on baryon resonances as well, but assume that baryon resonances
do no contribute to the strength of the mean field\footnote{We count
  members of baryon octet and decuplet as ground state baryons, and
  all other baryon states as resonances.}. We also use a more
realistic value for the mean-field parameter, $K=250\,{\rm MeV/fm}^3$,
and show the results for an exact mean-field calculation only.

In Fig.~\ref{fig:trace} we show our calculations for the trace
anomaly, $\epsilon-3P$, compared with the lattice results
\cite{Borsanyi:2013bia,Bazavov:2014pvz} at zero baryon chemical
potential, $\mu_B=0$. As expected the missing states have a clear
effect on the trace anomaly improving the fit to the lattice data:
These states lead to significant increase of the trace anomaly around
temperatures of $150$ MeV and higher compared to the HRG calculations
with only PDG states (PDG-HRG). The HRG calculations with repulsive
mean field are shown as dashed lines. The effect of the repulsive
interactions turns out to be relatively small, because the
contribution of baryons to the equation of state is small too. The
situation will be different at sufficiently large $\mu_B$, when the
contributions of mesons and baryons to the equation of state are
comparable.

\begin{figure}
 \hspace*{-7mm}
\includegraphics[width=8cm]{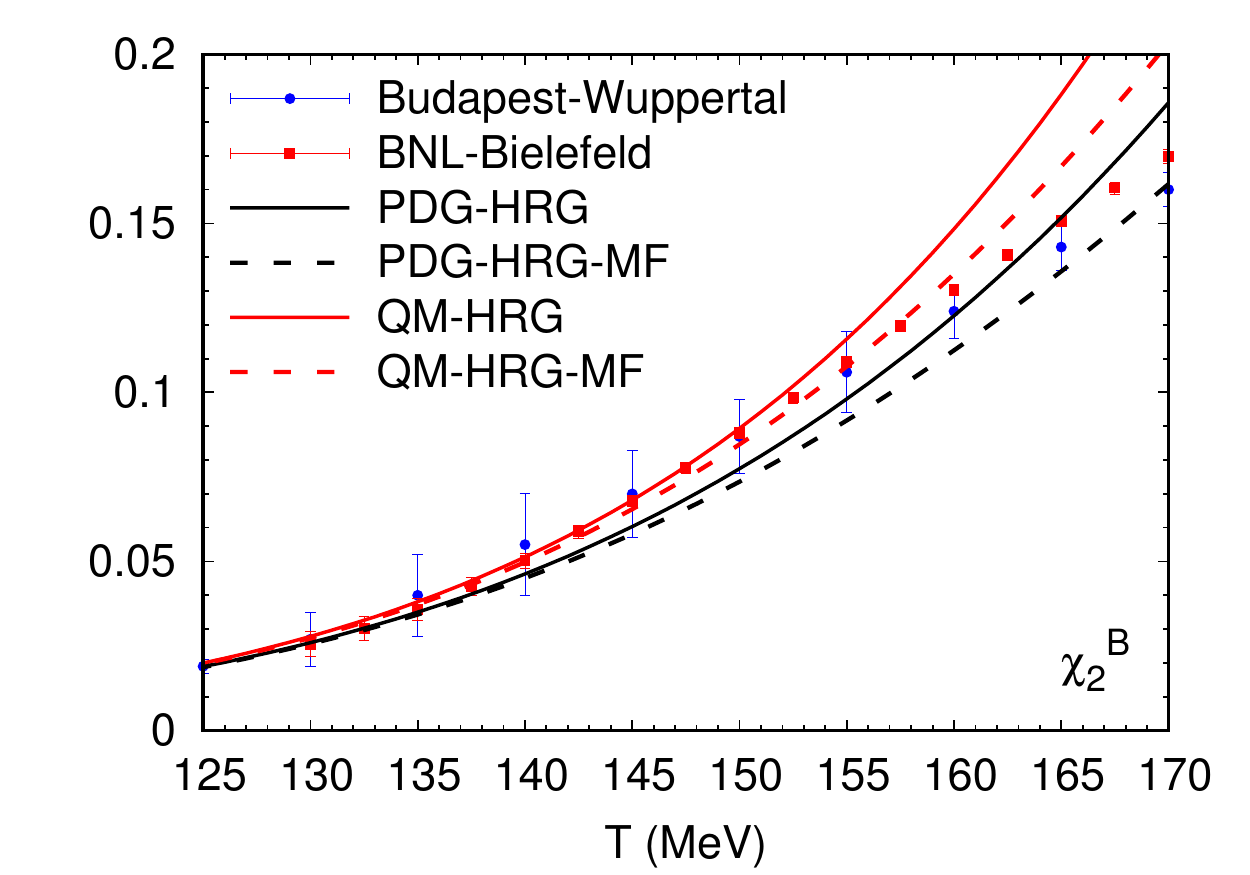}
\includegraphics[width=8cm]{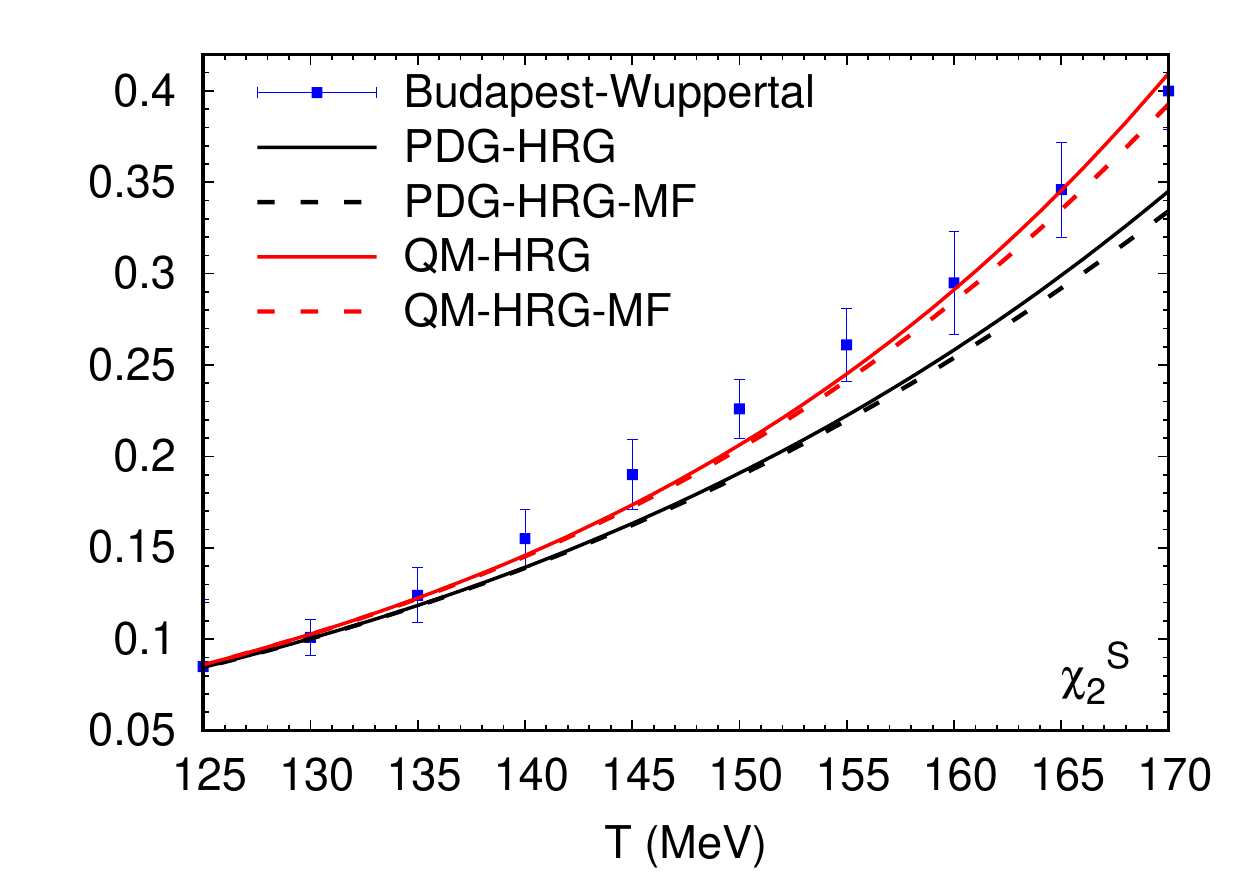}
\caption{The second order baryon number (left) and strangeness (right)
  fluctuations in HRG model with the Particle Data Group (PDG-HRG) and
  Quark Model (QM-HRG) lists of resonances compared to the lattice
  results \cite{Borsanyi:2014ewa,Bazavov:2017dus}.  The dashed lines
  correspond to HRG model with repulsive mean field.}
\label{fig:chi2B}
\end{figure}

In the left panel of Fig.~\ref{fig:chi2B} we show our results for
second order baryon number fluctuations, again for QM-HRG and PDG-HRG
with and without the effect of the repulsive mean field. The effect of
the missing states is clearly visible and improves the fit to lattice
data around 140--150 MeV temperature, but overshoots the lattice
results at higher temperatures. The repulsive mean field has an
opposite effect and is comparable in size. Therefore, it restores the
agreement with the lattice data. The second order strangeness
fluctuations are shown in the right panel of Fig.~\ref{fig:chi2B}. The
need for more resonance states to fit the data is now even clearer
than in the case of baryon number fluctuations, and even the full
quark model spectrum of states hardly reaches the lattice data. The
mean field, on the other hand, affects the strangeness fluctuations
less, since they are dominated by kaons and other strange mesons which
are not affected by baryonic interactions.

\begin{figure}
  \hspace*{-7mm}
\includegraphics[width=8cm]{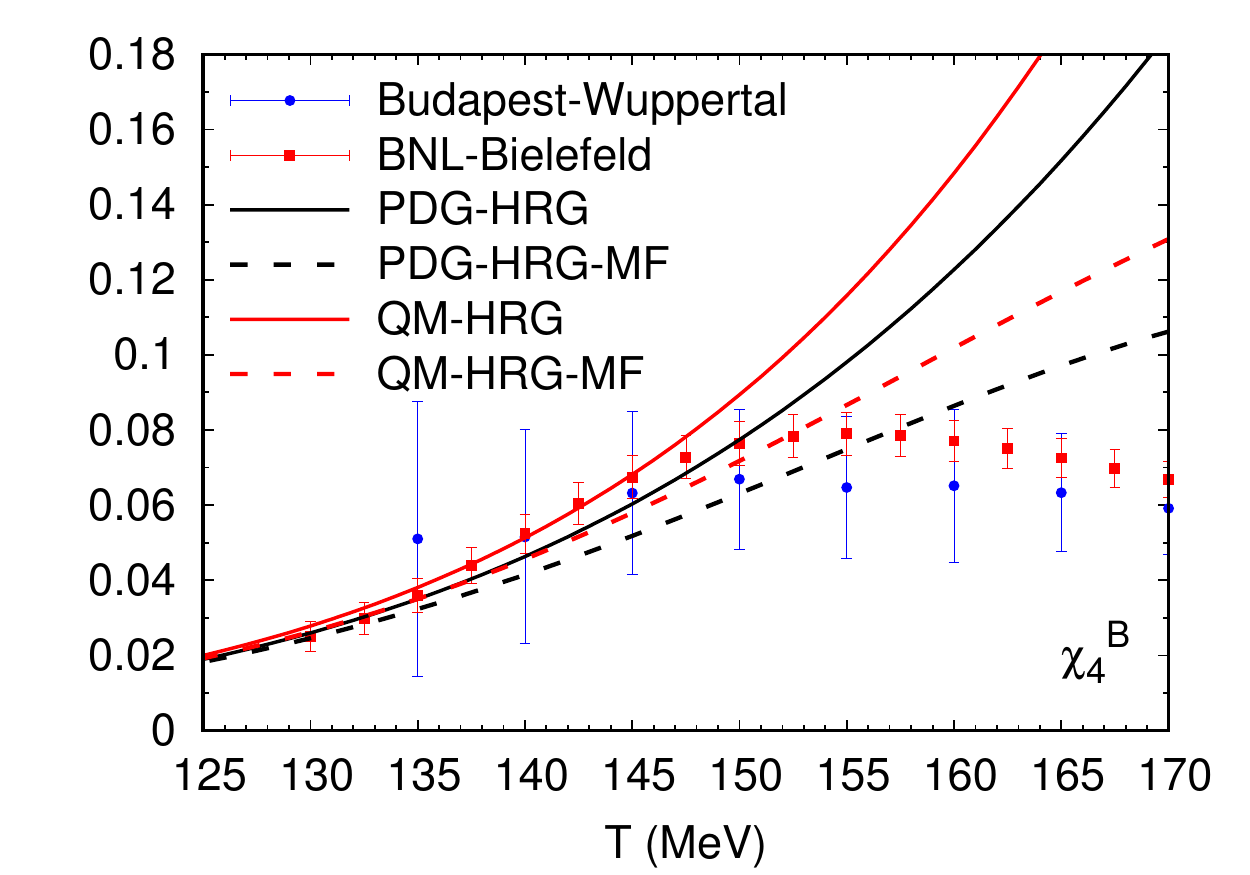}
\includegraphics[width=8cm]{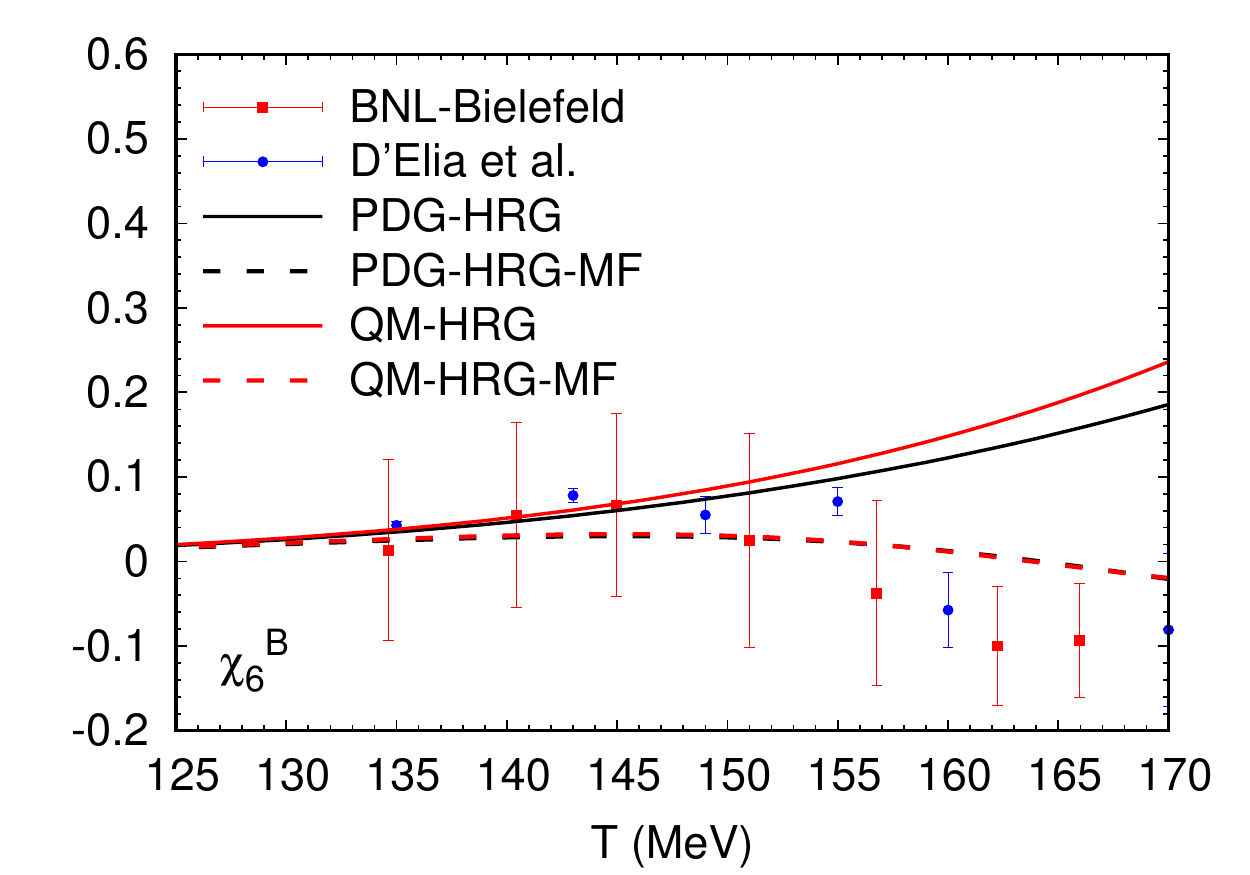}
\caption{The fourth order (left) and sixth order (right) baryon number
  fluctuations in HRG-QM and HRG-PDG models. The lattice results for
  $\chi_4^B$ are from Refs.~\cite{Borsanyi:2014ewa,Bazavov:2017dus},
  while the lattice results for $\chi_6^B$ are from
  \cite{DElia:2016jqh,Bazavov:2017dus}. The dashed lines correspond
  to HRG model with repulsive mean field.}
\label{fig:chi46B}
\end{figure}

Higher order baryon number fluctuations are more sensitive to the
effects of the repulsive interactions than the second order
fluctuations as shown in Fig.~\ref{fig:chi46B}. The QM-HRG model again
overpredicts the lattice results and lies significantly above the
PDG-HRG result. The repulsive interactions reduce the HRG prediction
and lead to reasonable agreement with the lattice data. Note that the
effects of the repulsive mean field are now larger than in the
previous analysis shown in Fig.~\ref{fig:old} even though we use
smaller value of $K$. This is because the resonances are also affected
by the repulsive mean field. Note also that when the effect of the
mean field is included, the sixth order baryon number fluctuation is
almost independent of the mass spectrum of resonances.

\section{Conclusion}
We studied the equation of state and fluctuations of baryon number and
strangeness within the HRG framework, where the effect of the
repulsive baryon-baryon interactions are included using the mean-field
approach. We also studied the effects of missing resonances on these
quantities. We have found that the missing states lead to significant
increase of thermodynamic quantities. In the case of baryon number
fluctuations missing states cause the HRG model to overshoot the
lattice results for $T \ge 150$ MeV. The repulsive mean field has the
opposite effect. The effects of repulsive interactions are small for
the trace anomaly and strangeness fluctuations, but are significant
for baryon number fluctuations, where they are needed to bring the HRG
calculations in agreement with the lattice results. This implies that
when extending the HRG model to calculate the equation of state at
non-zero baryon density the repulsive interactions have to be taken
into account.

\acknowledgments

This work was supported by National Science Center, Poland, under
grant Polonez DEC-2015/19/P/ST2/03333 receiving funding
from the European Union's Horizon 2020 research and innovation program
under the Marie Sk\l odowska-Curie grant agreement No 665778, and by\\
U.S.Department of Energy under Contract No.~DE-SC0012704.

\end{document}